\documentclass{jaa}

\usepackage{graphicx}
\usepackage[T1]{fontenc} 
\usepackage{subcaption} 

\begin{document}
\title{Long term Ultra-Violet Variability of Seyfert galaxies}


\author{N. Sukanya\textsuperscript{1}, C.S. Stalin\textsuperscript{2}, P. Joseph\textsuperscript{2}, S. Rakshit\textsuperscript{2,3}, D. Praveen\textsuperscript{4}, R. Damle\textsuperscript{1}}
\affilOne{\textsuperscript{1}Department of Physics, Bangalore University, Bangalore, 560 056, India\\}
\affilTwo{\textsuperscript{2}Indian Institute of Astrophysics, Block II,
 Koramangala, Bangalore, 560034, India.\\}
\affilThree{\textsuperscript{3}Department of Physics \& Astronomy, 
Seoul National University, Seoul, 08826, Republic of Korea\\}
\affilFour{\textsuperscript{4}Physics Department, Amrita School of Engineering, Bangalore 560 035, India\\}


\twocolumn[{

\maketitle

\corres{ramkrishnadamle@bub.ernet.in}

\begin{abstract}
Flux variability is one of the defining characteristics of Seyfert galaxies, 
a class of active galactic nuclei (AGN).  Though these variations
are observed over a wide range of wavelengths, results on their flux variability
characteristics in the ultra-violet (UV) band are very limited. We present
here the long term UV flux variability characteristics of a sample of fourteen
Seyfert galaxies using data from the International Ultraviolet Explorer
acquired between 1978 and 1995.
We found that all the sources showed flux variations with no statistically 
significant difference in the amplitude of UV flux variation between shorter and longer wavelengths.
Also, the flux variations between different near-UV
(NUV, 1850 $-$ 3300 \AA)
and far-UV (FUV, 1150 $-$ 2000 \AA) passbands in the rest frames of 
the objects are correlated with no time lag.
The data show indications of  (i) a mild negative correlation of UV variability 
with bolometric luminosity and (ii) weak positive correlation between 
UV variability and black hole mass.  At FUV, about 50\% of the sources
show a strong correlation between spectral indices and flux variations with a 
hardening
when brightening behaviour, while for the remaining sources the correlation
is moderate. In NUV, the sources do show a harder when brighter trend, however,
the correlation is either weak or moderate.

\end{abstract}


}]


\artcitid{\#\#\#\#}
\year{2018}
\pgrange{1--14}
\setcounter{page}{1}
\lp{14}

\section{Introduction}
Active galactic nuclei (AGN) with observed bolometric luminosities
of around 10$^{11}$ $-$ 10$^{14}$ L$_{\odot}$, and that include Seyfert galaxies
amongst its class, are believed to be powered by accretion
of matter onto super massive black holes residing at the centres of
galaxies (Lynden-Bell 1969, Rees 1984). According to the standard
picture, accretion leads to the formation of accretion disk that
emits black body radiation. The observed ultra-violet (UV)/optical radiation 
in AGN
is well represented by the superposition of several multi-temperature
black body components (Frank et al. 2002) and the observed big blue bump (BBB) in
AGN spectra is often attributed to the accretion disk.

AGN are known to show flux variations since their discovery and is now
considered one of their defining characteristics.
(Ulrich et al. 1997, Wagner \& Witzel 1995). Such flux variations
are seen on a range of
time scales from a fraction of hours to years
 over the complete
electromagnetic spectrum from low energy radio to high energy $\gamma$-rays
(Wagner \& Witzel 1995, Ulrich et al. 1997, Zhang et al. 2017, Giveon et al. 1999). In spite of having a wealth of monitoring data on
large samples of AGN with varying time resolutions through time domain
surveys as well as dedicated monitoring programs, the physical mechanisms
that cause AGN flux variations are still not well understood. Though
different physical processes contribute to the emission at different
wavebands, the UV-optical emission is believed to be emitted from an
optically thick and geometrically thin accretion disk (Frank et al.
2002). Therefore,
study of flux variations in the UV/optical bands can enable one to
understand the processes happening in the accretion disk of AGN in 
particular in the non-blazar category of AGN.

Earlier efforts on the study of UV variations in
AGN, were by Paltani \& Courvoisier (1994) who carried out a systematic
analysis of the flux variations in the UV of different classes of AGN
using data from the international Ultraviolet Explorer (IUE) covering the
period 1978 $-$ 1991. Also, UV variability of blazars has been
studied using IUE data (Edelson et al. 1991, Edelson 1992).
According to their analysis, blazars show stronger variability at shorter
wavelengths than at longer wavelengths.
Subsequent to the work of Paltani \& Courvoisier (1994), 
Welsh et al. (2011)
carried out a systematic study of the UV
variability of a large number of AGN using data from
the Galaxy Evolution Explorer (GALEX) data base. According to 
Welsh et al. (2011), the UV variability of
quasars is much more than their optical fluctuations and among the
UV bands, the variability observed in the far-UV (FUV; 1344$-$1786 \AA) band is larger than
the variability in the near-UV (NUV; 1771$-$2831 \AA) band with is also similar to that found
by Paltani \& Courvoisier (1994).  The analysis of
Paltani \& courvoisier (1994) failed to find any significant differences
between the UV properties of radio-loud and radio-quiet quasars prompting
the authors to suggest that the UV emission from AGN is independent
of the radio emission properties. 
Studies of optical flux variations in different categories of 
AGN indicate, blazars tend to show large amplitude and high duty cycle
of variability within a night compared to other radio-loud and
radio-quiet AGN (Stalin et al. 2004). On year like time scales, 
among Seyfert galaxies in the optical band, radio-loud sources
are more variable than their radio-quiet counterparts (Rakshit \& Stalin 2017).

Short time scale UV flux
variations of the order of 1000 to 10000 seconds were found in
the Seyfert 1 galaxy NGC 7469 by 
Welsh et al. (1998), using the
Faint Object
Spectrograph on the Hubble Space Telescope as well as
Fairall 9 (Lohfink et al. 2017).  Most of the studies on the UV flux variability
of AGN  (Sakata et al. 2011, Paltani \& Courvoisier 1994) either using
spectroscopy or broad band photometry indicate
that the UV flux variability characteristics of AGN
can be well described by accretion disk models.   Vanden Berk et al. (2004) 
based on two epochs of observations on a large number of quasars found a spectral
hardening of the UV continuum emission with increasing flux values. Similar results were also found 
by Wilhite et al. (2005) using spectroscopic observations.  Paltani \& Walter (1996) using 
IUE observations observed that the spectra of Seyfert galaxies vary with time and they become
flatter when the source brightens. To explain the observations, they 
proposed a two component model wherein the UV flux variations consist of a variable 
component with a constant spectral shape and a non-variable component from the 
small blue bump (SBB). 
Also, there
are reports that claim the constancy of UV spectral shape during flux 
variations of AGN (D. Alloin et al. 1995, Rodrigues-Pascual 
et al. 1997). As we have limited studies on
the UV flux variability characteristics of AGN both in long term as well
as short term, it is of great importance to expand the studies
on the already known UV flux variability nature of AGN to a larger sample
of sources, and having data for a longer duration of time than that 
analysed before by Paltani \& Walter (1996).  Towards this, we have carried out a statistical analysis of the
UV variability of a sample of Seyfert galaxies, a category of AGN
for which sufficient data is available  and focussed mainly on the
FUV (1150$-$2000 \AA) and
NUV (1850$-$3200 \AA) flux variations.
\section{Sample and Data}
Our sample of Seyfert 1 galaxies was taken from
Dunn et al. (2006) who have provided
continuum light curves in different wavebands for a  sample of 175 Seyfert
galaxies as
part of the Program in Extra Galactic Astronomy (PEGA)
\footnote{http://www.astro.gsu.edu/PEGA/IUE}. The data towards this compilation
were taken from the observations carried by 
IUE between the period 1978 to 1995. In this database, 
Dunn et al. (2006) have
provided continuum flux measurements at three line free regions in
the spectra of each of the Seyfert galaxies. In IUE spectra, the NUV and FUV cover the wavelength regions 1850$-$3200 \AA ~and 1150$-$2000 \AA ~respectively.
For most of the sources, flux measurements are available in three  NUV passbands
(2200, 2400 and 2740 \AA) with bin sizes of 50, 60 and 30 ~\AA  ~and  three
FUV passbands (1355, 1720 and 1810 \AA) with bin sizes of 30, 30 and 50 \AA.
 For this study, we have downloaded the
light curves for all the Seyfert galaxies that are available in
the PEGA database and we applied the
following conditions to select the light curves for 
further analysis:
\begin{enumerate}
\item The  sources must have data from the two cameras of IUE namely
the short wavelength prime (SWP) and long-wavelength prime (LWP).
\item The total number of points (that includes all the three continuum passbands
in FUV and NUV) must be larger than 50
\end{enumerate}

The above two conditions lead us to a final sample of 14 Seyfert galaxies
spanning the redshift range 0.002 $ < z < $ 0.07. Of the 14 selected Seyfert
galaxies, one galaxy (NGC 1068) belongs to the Seyfert 2 category (having
narrow permitted and forbidden lines), while the remaining 13 sources
belong to the Seyfert 1 category with broad permitted lines and narrow
forbidden lines. The details of the objects selected
for this study are given in Table~\ref{tab:details}. In this table, the
total in column 7 refers to the total number of photometric points for a 
source in all the six passbands together. The entries against $\lambda_1$, 
$\lambda_2$ and $\lambda_3$ in SWP and LWP columns refer to the central
wavelength used for the photometry and N$_{NUV}$ and N$_{FUV}$ give the 
number of points in each of the NUV and FUV passbands.
           
\begin{figure*}
\centering
\hbox{
\resizebox{5.5cm}{8cm}{\includegraphics{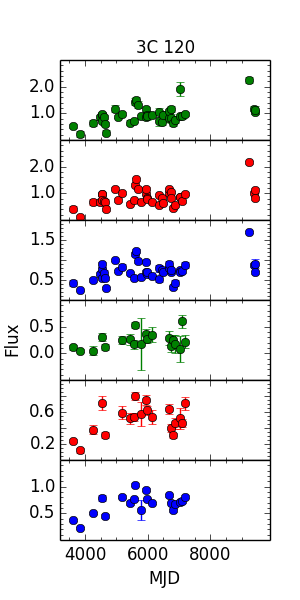}}
\hspace*{0.5cm}\resizebox{5.5cm}{8cm}{\includegraphics{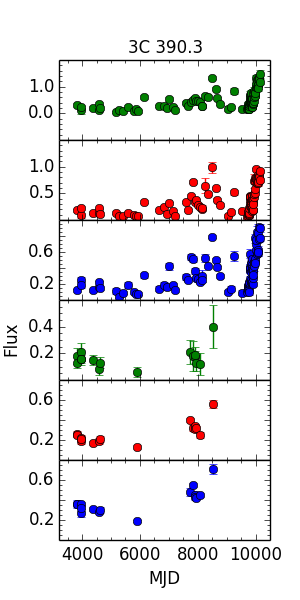}}
\hspace*{0.5cm}\resizebox{5.5cm}{8cm}{\includegraphics{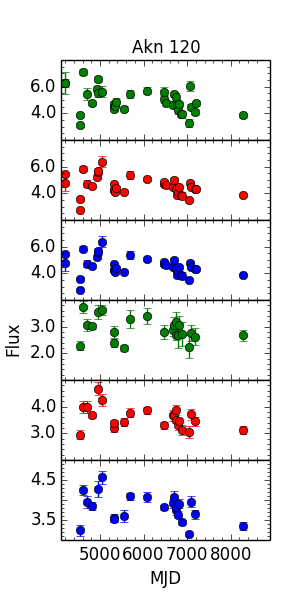}}
}
\caption{FUV and NUV light curves. The top three panels are for the FUV bands 
and the light curves in the bottom three panels are for NUV bands. For 3C 120, 
from top to bottom the wavelength of the light curves  are 1399 \AA, 1776 \AA,  
1869 \AA , 2272 \AA , 2479 \AA  ~and 2830 \AA. For 3C 390.3, the wavelength of 
the light curves from top to bottom are
1431 \AA, 1816 \AA, 1911 \AA, 2323 \AA, 2534 \AA ~and 2893 \AA. For Akn 120,
the light curves shown from top to bottom are in wavelengths of 
1390 \AA, 1775 \AA, 1868 \AA, 2271 \AA, 2477 \AA ~and 2828 \AA. } \label{Fig:1}
\end{figure*}

\begin{figure*}
\centering
\hbox{
\resizebox{5.5cm}{8cm}{\includegraphics{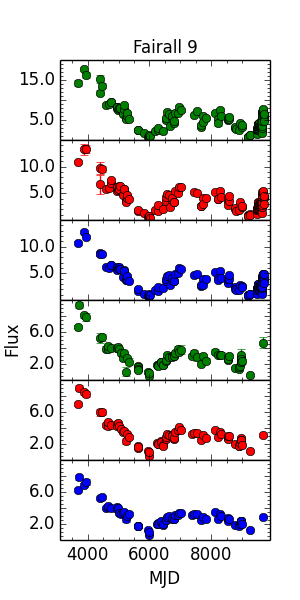}}
\hspace*{0.5cm}\resizebox{5.5cm}{8cm}{\includegraphics{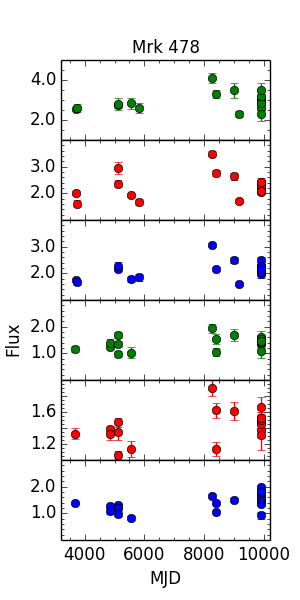}}
\hspace*{0.5cm}\resizebox{5.5cm}{8cm}{\includegraphics{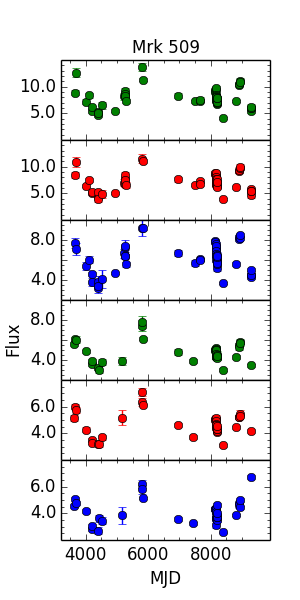}}
}
\caption{Light curves in increasing order of wavelengths from top
to bottom for Fairall 9 (left), Mrk 478 (middle) and
Mrk 509 (right). For Fairall 9, the wavelengths are 1418 \AA, 1800 \AA, 
1895 \AA, 2303 \AA, 2512 \AA ~and 2868 AA. ~For Mrk 478, the light curves
from top to bottom have wavelengths of 1462 \AA, 1855 \AA, 1953 \AA,
2373 \AA, 2589 \AA ~and 2956 \AA. For Mrk 509, from the top to the 
bottom panels, the light curves have the 
wavelengths of 1401 \AA, 1779 \AA, 1872 \AA, 2275 \AA, 2482 \AA ~and
2834 \AA.} \label{Fig:2}.
\end{figure*}

\begin{figure*}
\centering
\hbox{
\resizebox{5.5cm}{8cm}{\includegraphics{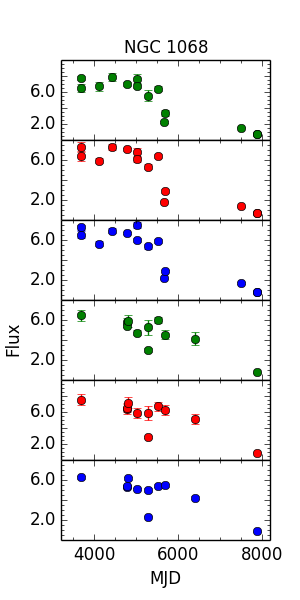}}
\hspace*{0.5cm}\resizebox{5.5cm}{8cm}{\includegraphics{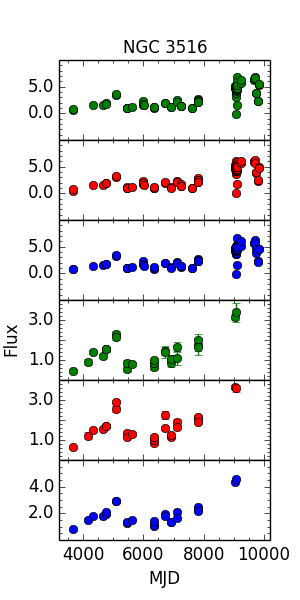}}
\hspace*{0.5cm}\resizebox{5.5cm}{8cm}{\includegraphics{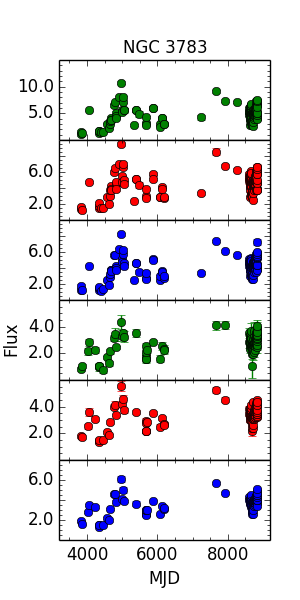}}
}
\caption{Light curves in FUV and NUV bands for the sources
NGC 1068 (left), NGC 3516 (middle) and NGC 3783 (right). For NGC 1068 the 
light curves from the top to the bottom panels are at wavelengths of 1360 \AA,
1726 \AA, 1816 \AA, 2208 \AA, 2409 \AA ~and 2750 \AA. For NGC 3516
the light curves are at increasing order of wavelengths from top to bottom
and have wavelength values of 1366 \AA, 1735 \AA, 1826 \AA, 2219 \AA, 
2421 \AA ~and 2764 \AA. For NGC 3783, the light curves from top to bottom panels
have wavelengths of 1368 \AA, 1736 \AA, 1827 \AA, 2221 \AA, 2423 \AA ~and 
2766 \AA. } \label{Fig:3}.
\end{figure*}

\begin{figure*}
\centering
\hbox{
\resizebox{5.5cm}{8cm}{\includegraphics{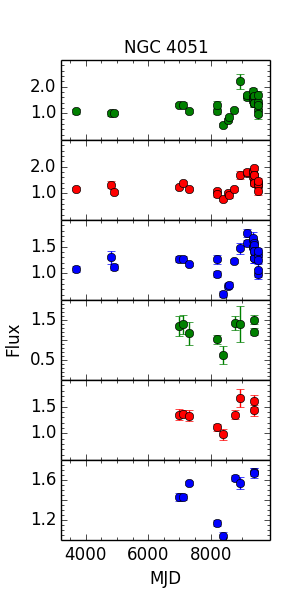}}
\hspace*{0.5cm}\resizebox{5.5cm}{8cm}{\includegraphics{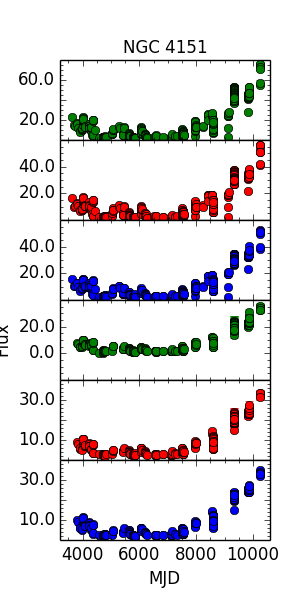}}
\hspace*{0.5cm}\resizebox{5.5cm}{8cm}{\includegraphics{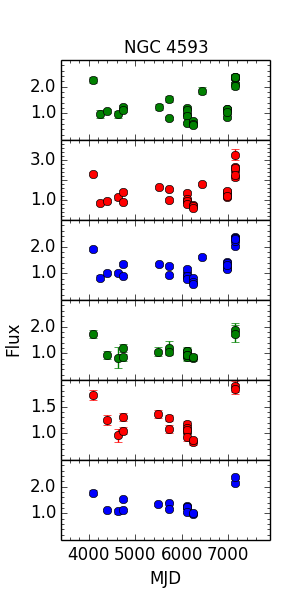}}
}
\caption{Light curves in FUV and NUV bands for  the 
sources NGC 4051, NGC 4151 and NGC 4593. For NGC 4051, 
the light curves in increasing order of wavelength from top to bottom 
are at wavelengths of 1358 \AA, 1724 \AA, 1814 \AA, 2205 \AA, 
2405 \AA ~and 2746 \AA. For NGC 4151, the light curves are at 
1359 \AA, 1725 \AA, 1816 \AA, 2207 \AA, 2407 \AA ~and 2749 \AA ~from the top
to the bottom panels. For NGC 4593, the shown light curves from top to bottom
are for wavelengths of 1367 \AA, 1735 \AA, 1826 \AA, 2219 \AA, 2421 \AA ~and
2764 \AA.} \label{Fig:4}.
\end{figure*}

\begin{figure*}
\centering
\hbox{
\resizebox{5.5cm}{8cm}{\includegraphics{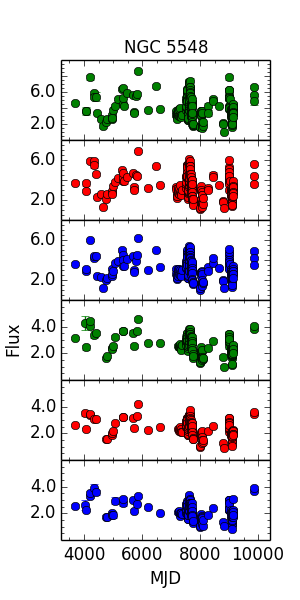}}
\hspace*{0.5cm}\resizebox{5.55cm}{8cm}{\includegraphics{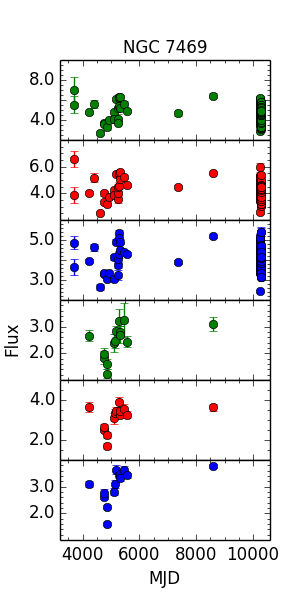}}
}
\caption{FUV and NUV light curves for the sources NGC 5548 
and NGC 7469. For NGC 5548, the wavelengths of the light curves from top to 
bottom are 1378 \AA, 1749 \AA, 1841 \AA, 2237 \AA, 2441 \AA ~and 2787 \AA.
For NGC 7469 the light curves shown from top to bottom have wavelengths of
1377 \AA, 1748 \AA, 1839 \AA, 2235 \AA, 2439 \AA ~and 2784 \AA.} \label{Fig:5}.
\end{figure*}

The observed flux values in all the six passbands were corrected for
galactic extinction using the $A_V$ values taken from NED \footnote{ned.ipac.
caltech.edu} which uses Schlafly et al. (2011) and the extinction
law evaluated in the UV range using the formalism given by
Cardelli et al. (1989). The galactic extinction corrected flux values
were then subjected to further analysis. We note here that the measured
flux values were not corrected for  extinction due to the host galaxies 
of the sources. The light curves in all the FUV and NUV passbands for the
sources are given in Fig. \ref{Fig:1} through Fig. \ref{Fig:5}. In these
figures, the quoted wavelengths are in the observed frame of the sources.
The present sample analysed here has some overlap with that reported
by Paltani \& Walter (1996). The sample analysed by Paltani \& Walter (1996)
has 15 sources that includes Seyfert galaxies, radio-loud as well as 
radio-quiet quasars. Their analysis was based on IUE data collected upto 1991.
Our sample analysed here contains 14 Seyfert galaxies using data from IUE upto 1995.
Though there are 10 sources in common to the sample reported here and that of
Paltani \& Walter (1996), the data analysed here is more extended in terms of the 
number of epochs and the total duration (1978$-$1995) compared to Paltani \& Walter (1996)
who have analysed data from IUE until 1991.
\subsection{Flux variability}
For all the 14 sources selected based on the criteria outlined 
above, we carried out analysis to characterise their variability.
This was done by calculating their normalized excess variance
defined in Vaughan et al. (2003)  as

\begin{equation}
F_{var} = \sqrt{\frac{S^2 - \overline{\sigma_{err}^2}} {\overline{x}^2}}
\end{equation}

where $S^2$ and $\sigma_{err}^2$ are the sample variance and average
error defined as

\begin{equation}
S^2 = \frac{1}{N-1} \sum_{i}(x_i - \overline{x})^2
\end{equation}

\begin{equation}
\overline{\sigma_{err}^2} = \frac{1}{N} \sum_{i=1}^N \sigma_{err,i}^2
\end{equation}

The error in $F_{var}$ was calculated again using Vaughan et al. (2003) and
is defined as
\begin{equation}\label{eq:ferr}
\sigma_{F_{var}} = \sqrt{\left(\sqrt{\frac{1}{2N}}\frac{\overline{\sigma_{err}^2}}{\overline{x}^2 F_{var}}\right)^2  + \left(\sqrt{\frac{\overline{\sigma_{err}^2}}{N}}\frac{1}{\overline{x}}\right)^2}
\end{equation}

\begin{table*}
\scriptsize
\caption{Details of the objects studied for variability. The M$_{BH}$ values are in solar units, and were taken from
the data base of Bentz \& Katz (2015) except for the sources NGC 1068 and Mrk 488 where
it was taken from  Greenhill \& Gwinn (1997) and 
Wang \& Lu (2001) respectively.}\label{tab:details}
\begin{tabular}{lccclccllcccccc} \hline
Name& RA(2000) &Dec (2000)& Redshift & log (M$_{BH}$)    & L$_{bol}$ & N$_{Total}$ & \multicolumn{4}{c}{SWP} & \multicolumn{4}{c}{LWP} \\ \cline{8-15}
   &   &   &   &  & (erg/sec)    & & $\lambda_1$  & $\lambda_2$       & $\lambda_3$ & N$_{FUV}$  & $\lambda_1$  & $\lambda_2$   & $\lambda_3$  & N$_{NUV}$ \\ \hline
Fairall 9 &  01:23:45.8 &  -58:48:20.5 &  0.047  & 8.299 & 44.78 &  693 & 1418 & 1800 & 1895 & 156 & 2303 & 2512 & 2868 &  75\\
NGC 1068  &  02:42:40.7 &  -00:00:47.8 &  0.004  & 7.176 & 44.60 &   75 & 1360 & 1726 & 1816 & 14  & 2208 & 2409 & 2750 &  11 \\
3C 120    &  04:33:11.1 &  +05:21:15.6 &  0.033  & 7.745 & 43.98 &  186 & 1399 & 1776 & 1869 & 42  & 2272 & 2479 & 2830 &  20 \\
Akn 120   &  05:16:11.4 &  -00:08:59.4 &  0.032  & 8.068 & 44.40 &  177 & 1398 & 1775 & 1868 & 36  & 2271 & 2471 & 2828 &  23 \\
NGC 3516  &  11:06:47.5 &  +72:34:06.9 &  0.009  & 7.395 & 43.89 &  330 & 1366 & 1735 & 1826 & 85  & 2219 & 2421 & 2764 &  25 \\
NGC 3783  &  11:39:01.8 &  -37:44:18.7 &  0.010  & 7.371 & 43.57 &  834 & 1368 & 1736 & 1827 & 164 & 2221 & 2423 & 2726 & 114\\
NGC 4051  &  12:03:09.6 &  +44:31:52.8 &  0.002  & 6.130 & 42.38 &  117 & 1358 & 1724 & 1814 & 30  & 2205 & 2405 & 2746 &  9 \\
NGC 4151  &  12:10:32.6 &  +39:24:20.6 &  0.003  & 7.555 & 43.30 & 2904 & 1359 & 1725 & 1816 & 542 & 2207 & 2407 & 2749 & 426\\
NGC 4593  &  12:39:39.4 &  -05:20:39.3 &  0.009  & 6.882 & 43.45 &  138 & 1367 & 1735 & 1826 & 29  & 2219 & 2421 & 2764 & 17 \\
NGC 5548  &  14:17:59.5 &  +25:08:12.4 &  0.017  & 7.718 & 43.92 & 1104 & 1378 & 1749 & 1841 & 214 & 2237 & 2441 & 2787 & 154\\
Mrk 478   &  14:42:07.5 &  +35:26:22.9 &  0.079  & 7.330 & 44.95 &  117 & 1462 & 1855 & 1953 & 19  & 2373 & 2589 & 2956 &  20 \\
3C 390.3  &  18:42:09.0 &  +79:46:17.1 &  0.056  & 8.638 & 44.32 &  330 & 1431 & 1816 & 1911 & 94  & 2323 & 2534 & 2893 &  16 \\
Mrk 509   &  20:44:09.7 &  -10:43:04.5 &  0.034  & 8.049 & 44.78 &  288 & 1401 & 1779 & 1872 & 55  & 2275 & 2482 & 2834 &  41 \\
NGC 7469  &  23:03:15.6 &  +08:52:26.4 &  0.016  & 6.956 & 44.42 &  750 & 1377 & 1748 & 1839 & 236 & 2235 & 2439 & 2784 & 14 \\
\hline
\end{tabular}
\end{table*}

\begin{table*}
\scriptsize
\caption{Results of the analysis of variability. The entries in columns 2, 3 and
4 are for the FUV bands, while the entries in columns 5,6 and 7 are for the 
NUV bands. Columns 8 and 9 give the mean $\alpha$ in SWP and LWP, and the last column 
gives the mean $\alpha$ estimated using IUE data covering the range of 1150 - 3200 \AA ~and taken 
from Paltani \& Walter (1996)}\label{tab:fvar_values}
\begin{tabular}{lcccccclll} \hline
Name      & \multicolumn{6}{c}{F$_{var} \pm \sigma_{F_{var}}$} & $\overline{\alpha}$(SWP) & $\overline{\alpha}$(LWP)  & $\overline{\alpha}$               \\ \cline{2-7}
          & \multicolumn{3}{c}{SWP}  & \multicolumn{3}{c}{LWP} &                 &                  &                 \\ \cline{2-7}
          &  $\lambda_1$       &  $\lambda_2$      & $\lambda_3$       & $\lambda_1$       &  $\lambda_2$      &  $\lambda_3$       &    &   &  \\ \hline
Fairall 9 &  0.586 $\pm$ 0.012 & 0.574 $\pm$ 0.012 & 0.562 $\pm$ 0.006 & 0.515 $\pm$ 0.022 & 0.499 $\pm$ 0.012 & 0.440 $\pm$ 0.009 & 0.92 $\pm$ 0.04 &  1.68 $\pm$ 0.05 & 0.9 \\
NGC 1068  &  0.529 $\pm$ 0.052 & 0.548 $\pm$ 0.036 & 0.522 $\pm$ 0.028 & 0.335 $\pm$ 0.073 & 0.339 $\pm$ 0.086 & 0.351 $\pm$ 0.022 & 1.68 $\pm$ 0.10 &  1.28 $\pm$ 0.15 & --- \\
3C 120    &  0.321 $\pm$ 0.058 & 0.327 $\pm$ 0.030 & 0.297 $\pm$ 0.026 & 0.414 $\pm$ 0.157 & 0.336 $\pm$ 0.052 & 0.282 $\pm$ 0.017 & 0.80 $\pm$ 0.09 &  2.46 $\pm$ 0.20 & 1.9 \\
Akn 120   &  0.174 $\pm$ 0.035 & 0.154 $\pm$ 0.022 & 0.154 $\pm$ 0.022 & 0.109 $\pm$ 0.057 & 0.104 $\pm$ 0.029 & 0.086 $\pm$ 0.017 & 1.41 $\pm$ 0.04 &  1.60 $\pm$ 0.07 & 1.5 \\
NGC 3516  &  0.600 $\pm$ 0.013 & 0.596 $\pm$ 0.012 & 0.590 $\pm$ 0.008 & 0.495 $\pm$ 0.044 & 0.435 $\pm$ 0.019 & 0.461 $\pm$ 0.011 & 1.75 $\pm$ 0.03 &  2.81 $\pm$ 0.10 & 2.2 \\
NGC 3783  &  0.305 $\pm$ 0.013 & 0.278 $\pm$ 0.010 & 0.269 $\pm$ 0.008 & 0.225 $\pm$ 0.035 & 0.217 $\pm$ 0.014 & 0.221 $\pm$ 0.007 & 1.15 $\pm$ 0.02 &  1.67 $\pm$ 0.04 & 1.5 \\
NGC 4051  &  0.240 $\pm$ 0.021 & 0.215 $\pm$ 0.016 & 0.206 $\pm$ 0.012 & 0.109 $\pm$ 0.076 & 0.137 $\pm$ 0.031 & 0.151 $\pm$ 0.011 & 1.91 $\pm$ 0.11 &  2.52 $\pm$ 0.13 & --- \\
NGC 4151  &  0.749 $\pm$ 0.011 & 0.728 $\pm$ 0.008 & 0.708 $\pm$ 0.006 & 0.681 $\pm$ 0.030 & 0.629 $\pm$ 0.016 & 0.650 $\pm$ 0.006 & 0.99 $\pm$ 0.02 &  2.25 $\pm$ 0.03 & 1.2 \\
NGC 4593  &  0.440 $\pm$ 0.016 & 0.462 $\pm$ 0.018 & 0.407 $\pm$ 0.011 & 0.246 $\pm$ 0.048 & 0.254 $\pm$ 0.019 & 0.284 $\pm$ 0.012 & 2.03 $\pm$ 0.08 &  2.53 $\pm$ 0.07 & 2.0 \\
NGC 5548  &  0.362 $\pm$ 0.008 & 0.337 $\pm$ 0.006 & 0.315 $\pm$ 0.004 & 0.253 $\pm$ 0.008 & 0.262 $\pm$ 0.004 & 0.271 $\pm$ 0.006 & 1.22 $\pm$ 0.02 &  1.06 $\pm$ 0.04 & 1.3 \\
Mrk 478   &  0.119 $\pm$ 0.039 & 0.196 $\pm$ 0.018 & 0.153 $\pm$ 0.016 & 0.134 $\pm$ 0.032 & 0.119 $\pm$ 0.019 & 0.226 $\pm$ 0.012 & 1.00 $\pm$ 0.13 &  1.88 $\pm$ 0.08 & --- \\
3C 390.3  &  0.712 $\pm$ 0.027 & 0.780 $\pm$ 0.018 & 0.645 $\pm$ 0.014 & 0.144 $\pm$ 0.056 & 0.367 $\pm$ 0.016 & 0.309 $\pm$ 0.015 & 1.57 $\pm$ 0.15 &  4.43 $\pm$ 0.18 & --- \\
Mrk 509   &  0.234 $\pm$ 0.024 & 0.230 $\pm$ 0.027 & 0.226 $\pm$ 0.021 & 0.209 $\pm$ 0.024 & 0.183 $\pm$ 0.016 & 0.207 $\pm$ 0.018 & 1.13 $\pm$ 0.04 &  0.80 $\pm$ 0.04 & 1.2 \\
NGC 7469  &  0.215 $\pm$ 0.047 & 0.204 $\pm$ 0.029 & 0.183 $\pm$ 0.020 & 0.213 $\pm$ 0.068 & 0.189 $\pm$ 0.037 & 0.195 $\pm$ 0.021 & 1.31 $\pm$ 0.02 &  1.73 $\pm$ 0.16 & 1.4 \\
\hline
\end{tabular}
\end{table*}

\begin{table}
\caption{Average F$_{var}$ values for the different wavelength bands}
\label{tab:afvar} 
\begin{tabular}{cc} \hline
Mean wavelength & Mean F$_{var}$ \\ \hline
1389 $\pm$ 30   & 0.399 $\pm$ 0.198 \\
1762 $\pm$ 38   & 0.402 $\pm$ 0.203 \\
1855 $\pm$ 40   & 0.374 $\pm$ 0.188 \\
2255 $\pm$ 49   & 0.292 $\pm$ 0.168 \\
2460 $\pm$ 53   & 0.291 $\pm$ 0.148 \\
2806 $\pm$ 64   & 0.295 $\pm$ 0.139 \\ \hline
\end{tabular}
\end{table}

\begin{table*}
\small
\caption{Results of the linear least squares fit between the
variation of spectral indices and the fluxes at the
shortest wavelength in NUV and FUV bands. The columns r and P
are the correlation coefficient and probability respectively.}\label{tab:specvar} 
\begin{tabular}{lccclcllcccccc} \hline
Name      & \multicolumn{4}{c}{SWP}  & \multicolumn{4}{c}{LWP} \\ \cline{2-8}
          &  Slope  & Intercept  &  r  &  P   &  slope  & Intercept & r  & P \\ \hline
Fairall 9 &  -0.100 $\pm$ 0.012 & 1.425 $\pm$ 0.083 & 0.405   & 0.000 & -0.393 $\pm$ 0.033  & 2.965 $\pm$ 0.105 & 0.601 & 0.000   \\
NGC 1068  &   -0.065$\pm$ 0.038 & 2.188 $\pm$ 0.353 & 0.530   & 0.052 & 0.009 $\pm$ 0.078   & 1.075 $\pm$ 0.489 & -0.109 & 0.751  \\
3C 120    &  -0.178 $\pm$ 0.038 & 2.268 $\pm$ 0.326 & 0.455   & 0.044 & -0.622 $\pm$ 0.116  & 4.131 $\pm$ 0.345 & 0.760 & 0.002   \\
Akn 120   &  -0.131 $\pm$ 0.026 & 3.097 $\pm$ 0.301 & 0.530   & 0.001 & -0.291  $\pm$ 0.071 & 3.866 $\pm$ 0.556 & 0.653 & 0.001  \\
NGC 3516  &   -0.079 $\pm$ 0.007 & 2.157 $\pm$ 0.036 & 0.286  & 0.008 & 0.074 $\pm$ 0.026   & 2.577 $\pm$ 0.098 & -0.101 & 0.637   \\
NGC 3783  &  -0.088 $\pm$ 0.006 & 2.204  $\pm$ 0.068 & 0.683  & 0.000 & -0.262 $\pm$ 0.030  & 3.634 $\pm$ 0.217 & 0.329  & 0.000   \\
NGC 4051  &   -1.140 $\pm$ 0.223 & 3.653 $\pm$ 0.326 & 0.596  & 0.001 & -1.525 $\pm$ 0.873  & 4.472 $\pm$ 1.199 & 0.285  & 0.457   \\
NGC 4151  &   -0.019 $\pm$ 0.001 & 1.310 $\pm$ 0.027 & 0.720  & 0.000 & -0.058 $\pm$ 0.006  & 2.951 $\pm$ 0.090 & 0.380  & 0.000  \\
NGC 4593  &   -0.002 $\pm$ 0.007 & 2.302 $\pm$ 0.018 & 0.378  & 0.043 & -1.798 $\pm$ 0.626  & 4.939 $\pm$ 0.802 & 0.010  & 0.971   \\
NGC 5548  &   -0.234 $\pm$ 0.019 & 2.280 $\pm$ 0.085 & 0.547  & 0.000 & -0.547 $\pm$ 0.075  & 2.533 $\pm$ 0.250 & 0.179  & 0.050  \\
Mrk 478   &   -1.201 $\pm$ 0.621 & 4.474$\pm$ 1.896 & 0.181  & 0.518 & -5.171 $\pm$ 2.293  & 9.716 $\pm$ 3.427 & 0.177 & 0.483   \\
3C 390.3  &  -1.043 $\pm$ 0.142 & 2.160 $\pm$ 0.172 & 0.609   & 0.000 & -12.695 $\pm$ 4.974 & 7.483 $\pm$ 1.222 & 0.445 & 0.084   \\
Mrk 509   &  -0.038 $\pm$ 0.017 &  1.653 $\pm$ 0.191 & 0.097  & 0.485 & -0.318 $\pm$ 0.098  & 2.956 $\pm$ 0.722 & 0.388 & 0.031     \\
NGC 7469  &   -0.221 $\pm$ 0.009 & 2.982 $\pm$ 0.067 & 0.786  & 0.000 & -0.211 $\pm$ 0.093  & 2.846 $\pm$ 0.460 & 0.360  & 0.206  \\ \hline
\end{tabular}
\end{table*}

\begin{figure}
\centering
\hspace*{-0.2cm}\resizebox{9.5cm}{8cm}{\includegraphics{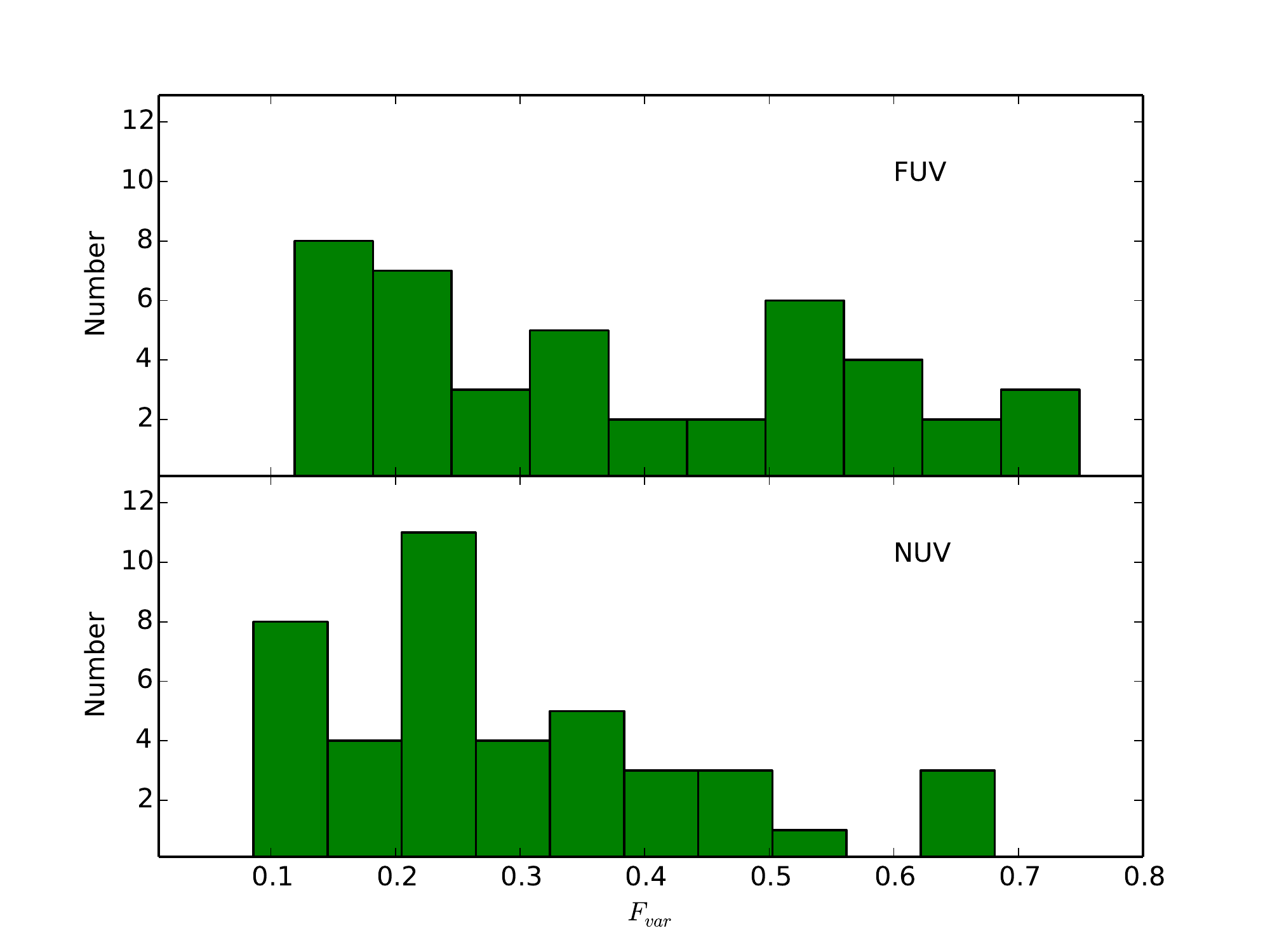}}
\caption{Distribution of $F_{var}$ values for the sources studied here in
FUV (top panel) and NUV (bottom panel) bands}.\label{Fig:6}
\end{figure}

\begin{figure}
\centering
\hspace*{-0.2cm}\resizebox{9.5cm}{8cm}{\includegraphics{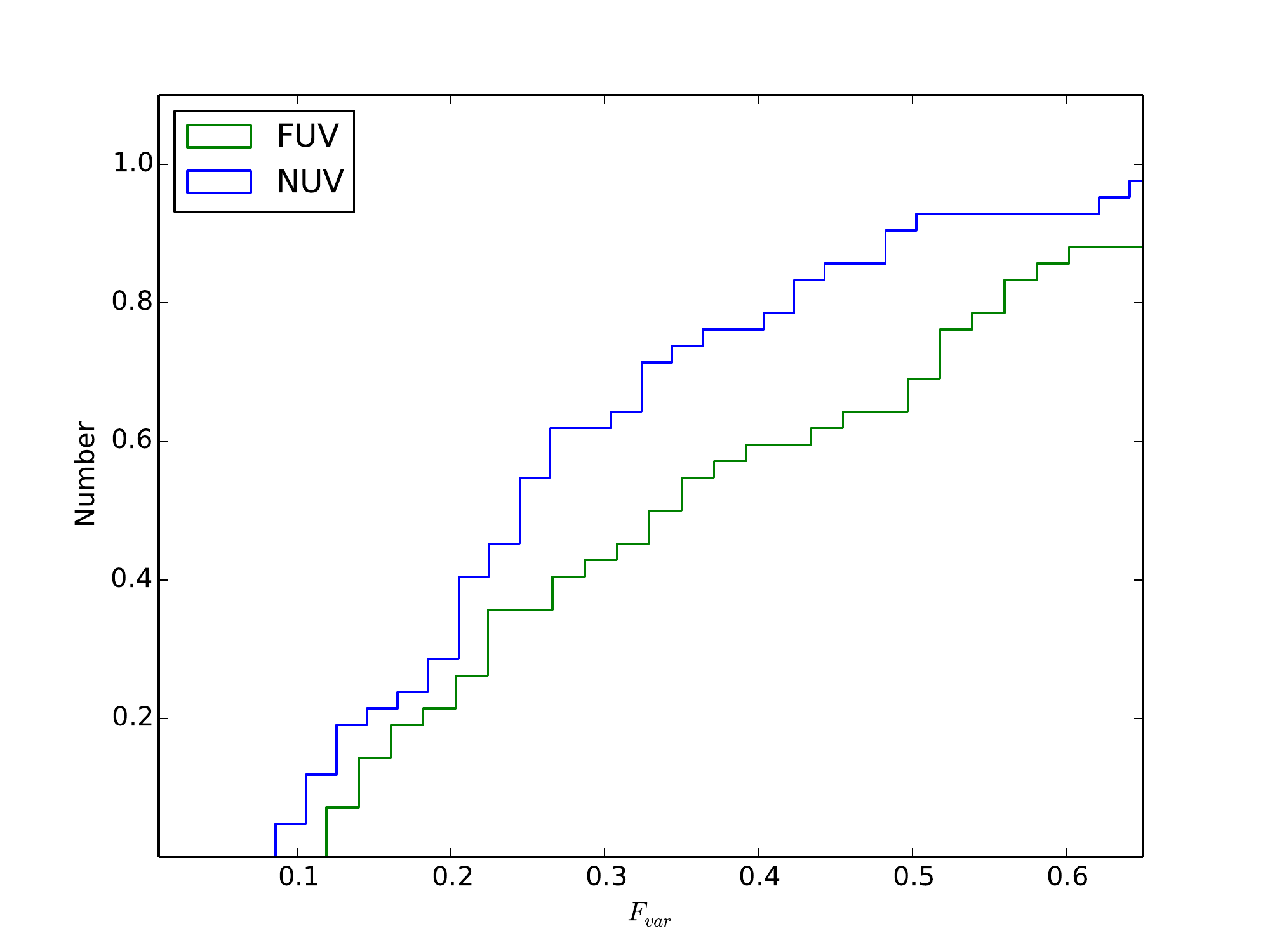}}
\caption{Cumulative distribution of the values of F$_{var}$ in FUV (in green) 
and NUV (in blue) bands}.\label{Fig:7}
\end{figure}

 Majority of the sources in our sample, have overlapping coverage in FUV
and NUV passbands except for five sources, namely, 3C 120, 3C 390.3, 
NGC 3516, NGC 7469 and NGC 4051. This is evident in the light curves
of these sources shown in Figs. 1, 3 and 5. Because of this, for calculating 
F$_{var}$, we have considered only those duration of 
the light curves that have overlapping coverage in both FUV and NUV passbands.
The calculated F$_{var}$ values for all the sources in each of the six
continuum passbands  are given in Table~\ref{tab:fvar_values}.
A source is considered variable if its F$_{var}$ is greater than zero,
and it
is significant at the one sigma level.  In all instances in our sample, F$_{var}$
is many times greater than their associated errors except in four cases
where it is less than three times their associated errors. Among these
four too, in two cases F$_{var}$ is more than two times their associated errors
and in the remaining two cases it is between one and two sigma. 
We therefore argue that all the sources in the sample analysed here are highly variable
in all the six passbands, except in four instances where the 
variability is less significant. The mean F$_{var}$ values in all the six passbands
for all the sources studied here is shown in Table~\ref{tab:afvar}.  There is an indication that the variations
at the shorter wavelengths are larger than those at the longer wavelengths,
but the larger error bars preclude us to draw any firm conclusion on the
differences in variability between
different wavelengths. Clubbing the F$_{var}$ values in the three SWP passbands
together
as SWP and the three LWP passbands together as LWP, we obtained simple average values of F$_{var}$
as 0.392 $\pm$ 0.196 and 0.293 $\pm$ 0.152 for SWP and LWP respectively. 
The total number of F$_{var}$ values are thus 42, in each of the SWP and LWP passbands. The
distributions of F$_{var}$ for SWP and LWP and their cumulative
distributions are given in Fig. \ref{Fig:6} and Fig. \ref{Fig:7} 
respectively.  The average
value of F$_{var}$ is larger in SWP than in LWP, however, as the
error bars are larger we carried out a two sample KS test.
The null hypothesis that was tested was that the two independent
F$_{var}$ values pertaining to SWP and LWP were drawn from
the same distribution. This null hypothesis was accepted as D was lesser
than the critical value of D (D$_{crit})$. We obtained values of 0.286 and 0.356 for D and D$_{crit}$
respectively for a significance level of 0.01.  This statistically points to no difference in the F$_{var}$
values between SWP and LWP bands. Available studies do indicate that in UV, 
AGN show wavelength dependent variability with the shorter wavelengths showing
large amplitude of variability compared to the longer wavelengths
(Sakata et al. 2011, Vanden Berk et al. 2004, Welsh 
et al. 2011). Data analysed here do indicate
that variations at the shorter wavelengths are larger than that at longer
wavelengths, however, due to the quality of the data, the error bars are too
large to draw any conclusion on variation of amplitude of 
variability with wavelength.

\begin{figure}
\centering
\hspace*{-0.2cm}\resizebox{9.5cm}{8cm}{\includegraphics{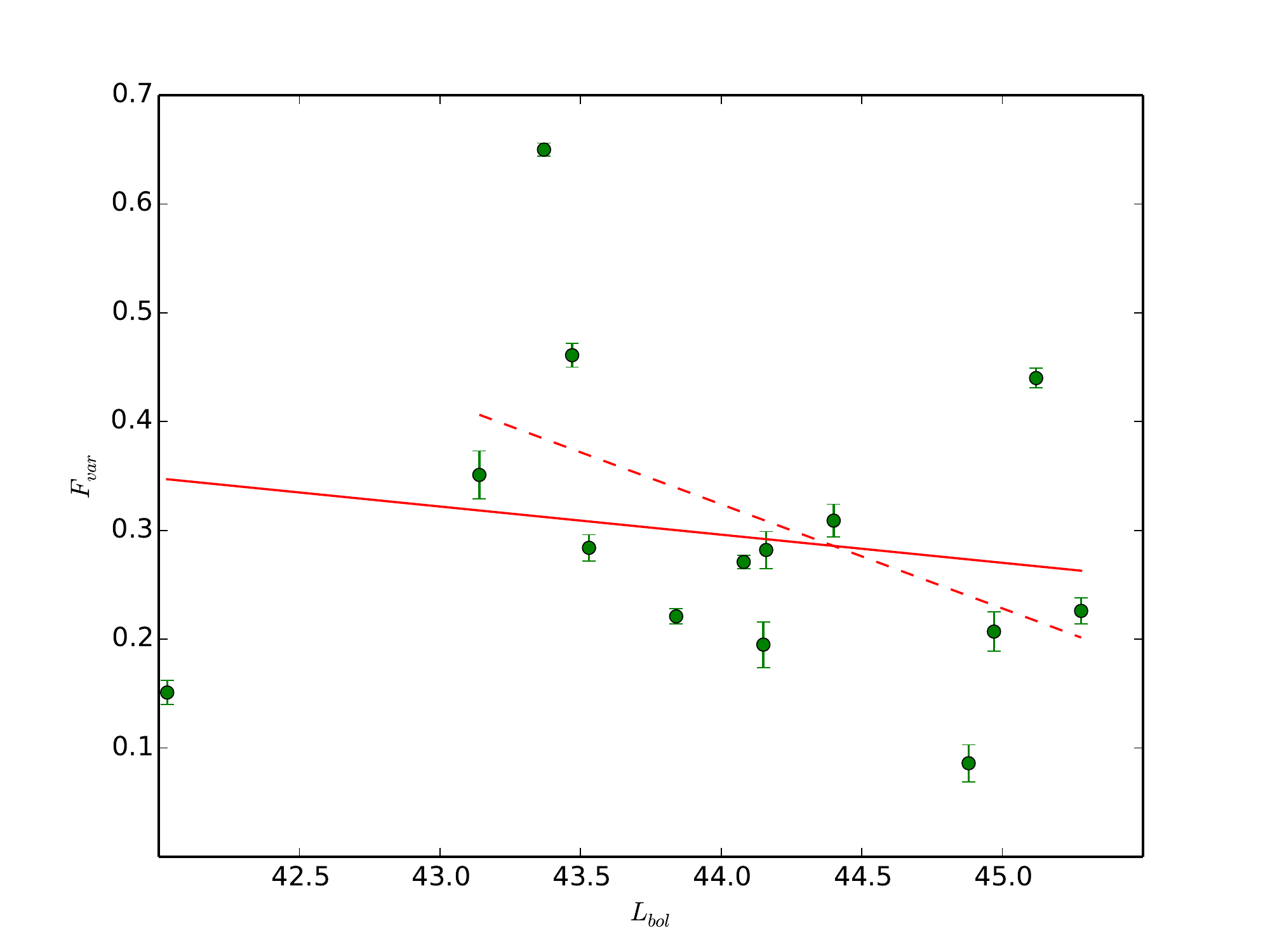}}
\caption{ Variation of F$_{var}$ with bolometric luminosity. Linear least
squares fit to the data are shown for the complete data (solid line) and
for the data set excluding the lowest luminosity source in our sample (dashed
line)}. \label{Fig:8}
\end{figure}

\begin{figure}
\centering
\hspace*{-0.2cm}\resizebox{9.5cm}{8cm}{\includegraphics{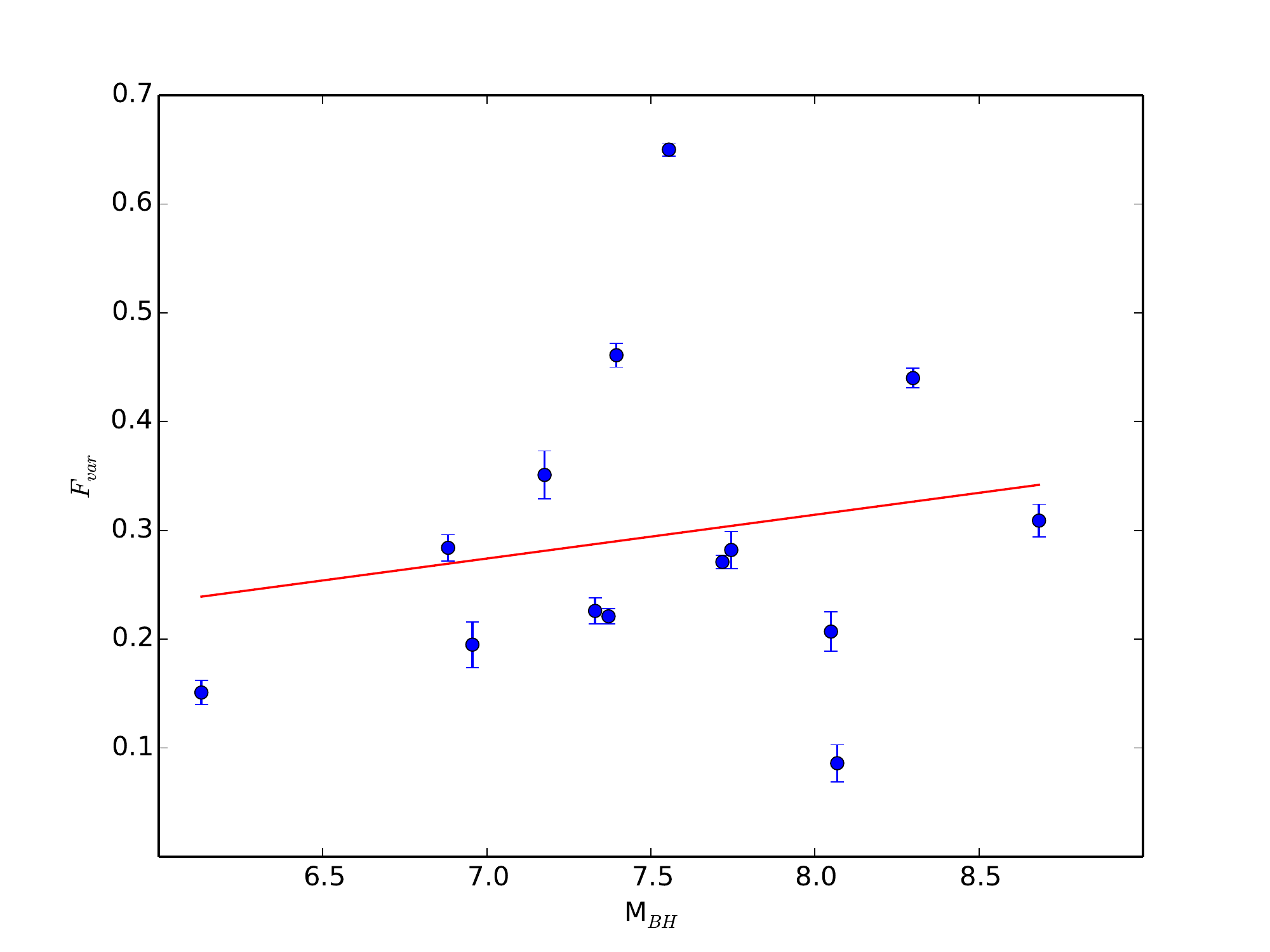}}
\caption{Plot of F$_{var}$ against black hole mass. The solid
line is the linear least squares fit to the data.} \label{Fig:9}.
\end{figure}

\section{Correlation between variability and other physical properties}
\subsection{F$_{var}$ and  L$_{bol}$}
To find for the presence of any correlation between F$_{var}$ and
bolometric luminosity (L$_{bol}$), we plot in Fig. \ref{Fig:8} the variation
of F$_{var}$ with L$_{bol}$ . The F$_{var}$ values used in this correlation 
analysis is for the NUV band for the passband 2806 $\pm$ 64 \AA.
We used the relation $L_{bol} = 13.2 \times L_{V}$
given by Elvis et al. (1994). Here, $L_V$ is the luminosity
in the V-band which was derived using the V-band magnitude
of the sources taken from SIMBAD\footnote{http://simbad.u-strasbg.fr/simbad/}, the zero-points taken from Bessel
(1979) and the luminosity distance taken from NED\footnote{http://www.astro.ucla.edu/$\sim$wright/CosmoCalc.html}.
Using all the F$_{var}$ values, we found indication of 
no correlation between F$_{var}$ and $L_{bol}$  with a low correlation
coefficient of $-$0.08 and a probability of no correlation of P = 0.79.
This trend for no correlation between F$_{var}$ and L$_{bol}$ is due to one
low luminosity source NGC 4051. Neglecting this source and doing a linear least
squares fit to the data gave evidence for a mild negative correlation
between F$_{var}$ and L$_{bol}$. The linear least square fit is shown 
as a dashed line in Fig. \ref{Fig:8}. Correlation analysis indicates 
a  mild negative correlation with a correlation coefficient of $-$0.39 with
a probability of no correlation of P = 0.19.  This is in agreement with
what is known in literature.
Using IUE data,
Paltani \& Courvoisier (1997) found an anti-correlation between quasar
variability and luminosity with high luminosity quasars showing low amplitude
of variability. This anti-correlation is also seen in the optical bands
(Vanden Berk et al. 2004, Meusinger \& Weiss 2013). Analysing large sample of quasars for 
UV variability using data from GALEX Welsh et al. (2011) found 
two different correlations between variability and luminosity. For time lags greater
than 100 days, variability is negatively correlated with luminosity, while for time lags
lesser than 100 days, variability is positively correlated with luminosity. The data
analysed here too reveal a negative correlation  between UV variability and 
luminosity. However, quality UV data (with similar time resolution and uniform coverage
in both FUV and NUV) on a larger sample of sources are needed 
to firmly  establish this finding.

\subsection{F$_{var}$ and M$_{BH}$}
Correlation between optical variability and black hole (BH) mass has been widely 
studied in the optical with no clear consensus. 
From an analysis of the long term optical variability of quasars, 
Wold et al. (2007) found a correlation between variability and BH mass with
sources with large BH mass showing larger amplitude of variability. 
Such a correlation was also noticed by Wilhite et al. (2008), however,
Meusinger \& Weiss (2013) and Zuo et al. (2012) could not find any correlation between
optical variability and BH mass. From the data set analysed here, we 
looked for the existence of any correlation between UV variability and 
BH mass. In Fig. \ref{Fig:9} we show the correlation between 
F$_{var}$ and M$_{BH}$ where we found hint for a positive correlation between
F$_{var}$ and M$_{BH}$. Correlation analysis gave a Pearson rank correlation
coefficient of 0.18 with a probability for no correlation of 0.54. The F$_{var}$
values used in this correlation analysis too is in NUV for the passband
2806 $\pm$ 64 \AA.

\section{Spectral variability}
To know the spectral variability nature of the sources studied here, we examined
the change  in the spectral index relative to the flux of the sources.
The optical to UV continuum slope of an AGN can be well represented as
a power law, $F_{\nu} \propto \nu^{-\alpha}$, where $F_{\nu}$ is the observed
flux density and $\alpha$ is the spectral index. For each of the sources
studied here, we have observations in six UV passbands. We therefore
calculated the spectral index by fitting a power law of the form
\begin{equation}
F_{\lambda} \propto \lambda^{\alpha-2}
\end{equation}

\begin{figure*}
\centering
\hspace*{-0.2cm}\includegraphics[scale=0.5]{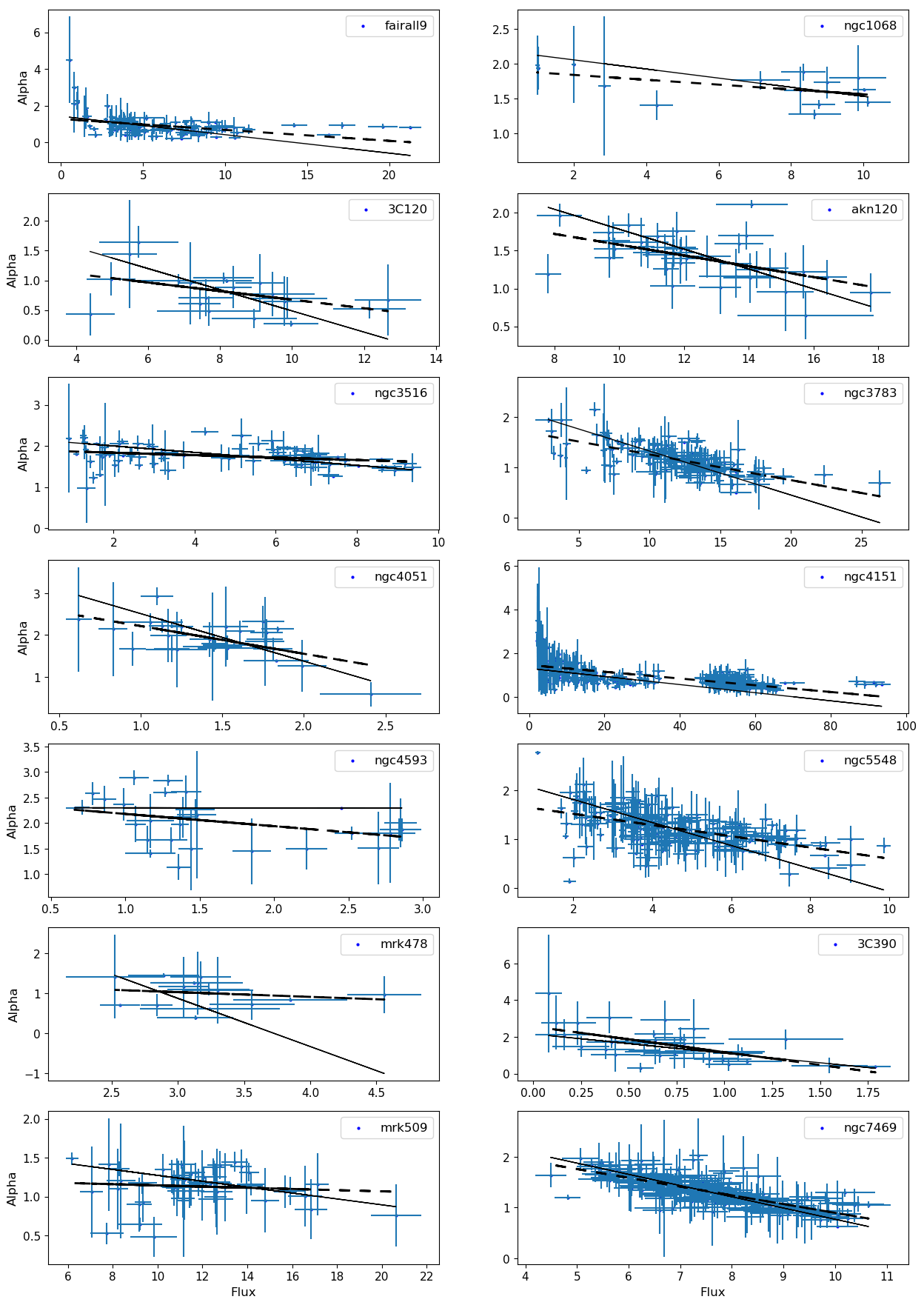}
\caption{Variation of spectral index with flux in the shortest
wavelength in FUV. The solid lines are the linear least squares fit to the
data that take into account the errors in both $\alpha$ and fluxes, while the dashed
lines are the un-weighted linear least squares fit to the data.} \label{Fig:10}.
\end{figure*}

\begin{figure*}
\centering
\hspace*{-0.2cm}\includegraphics[scale=0.5]{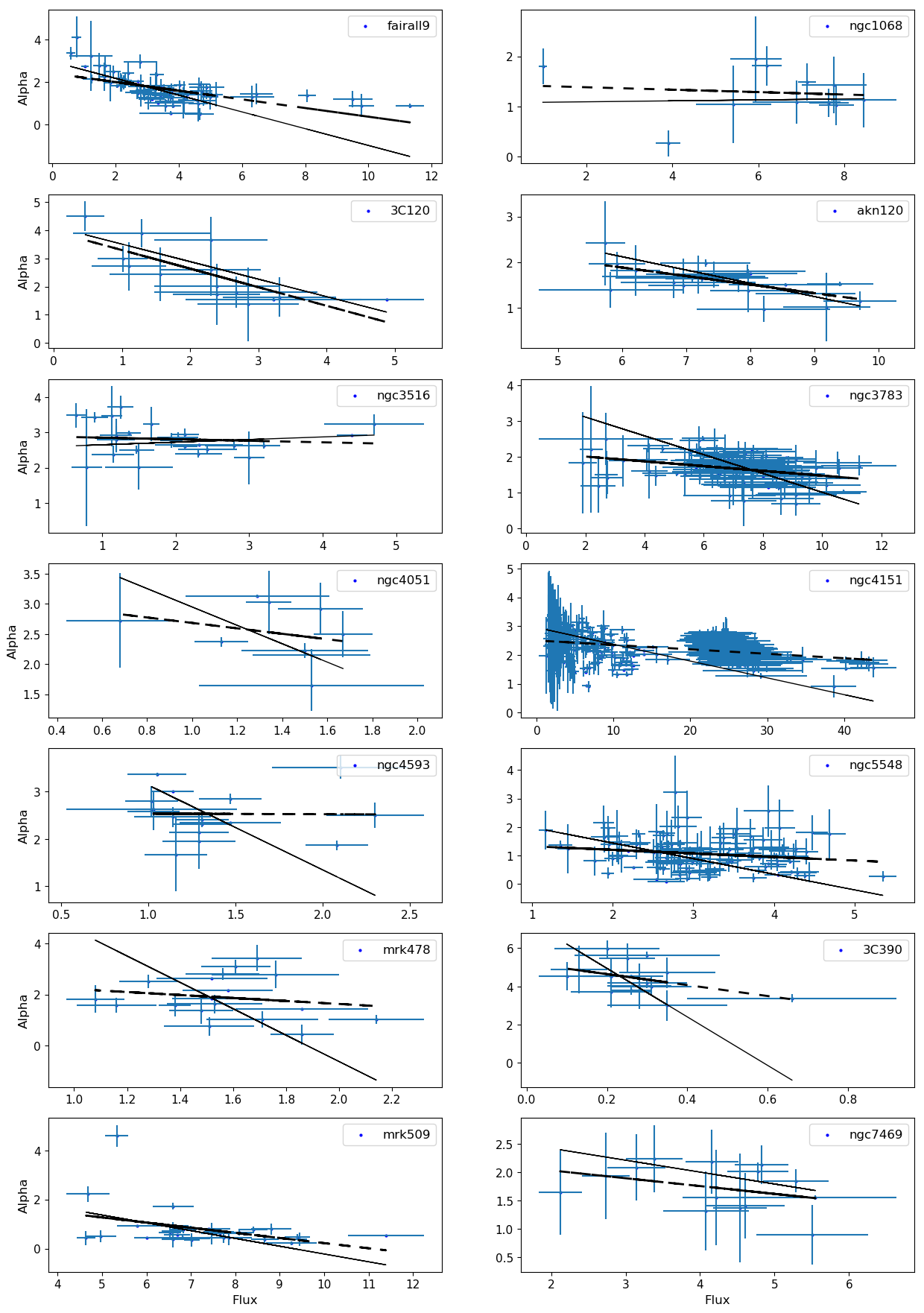}
\caption{Variation of spectral index with flux in the NUV band. Linear
least squares fit to the data that takes into account the errors in both
the spectral indices and flux values are shown as solid lines. The un-weighted 
linear least squares fit to the data are shown with dashed lines.} \label{Fig:11}
\end{figure*}

For NUV, $\alpha$ was determined using the above power law fit to three
measurements and for FUV again three measurements were used to derive
$\alpha$.  The variation of $\alpha$ thus deduced against
the flux of the sources in both SWP and LWP are shown in 
Fig.\ref{Fig:10} and Fig.\ref{Fig:11}
respectively.
For SWP, we selected the shortest wavelength of three and for LWP too, we
selected the shortest wavelength of the three observations. This is 
only for the purpose of defining the flux values. For the
analysis of correlation between $\alpha$ and flux, we considered
only those points where the error in $\alpha$ and flux values are lower than the 
associated values of $\alpha$ and fluxes. The data were fit with
a straight line by (i) assigning equal weight to all the points and (ii) taking
into account the errors in both $\alpha$ and flux values. The un-weighted linear
least squares fits are shown by dashed lines in Fig. 10 and Fig. 11, while the 
weighted linear least squares fits are shown by solid lines. From weighted linear
least squares fit to the data we find that, for most of the
sources, their spectra do not show any significant changes  during the
flux variations, however, for few sources, we found clear evidence
of a hardening of the spectra with increase in flux. For some
sources, we see structures in the variation of $\alpha$ with flux. The spectrum is
found to harden with increasing flux, however, limited
to certain moderate flux values, beyond which the spectrum is nearly steady showing 
no change with flux. This is seen in the sources NGC 3783, NGC 4151, NGC 4593 and Fairall 9 
in SWP. In LWP, this is evident in the sources NGC 4151 and Fairall 9. 
The results of the linear
least squares fit to the variation in $\alpha$ with flux is shown in
Table~\ref{tab:specvar} . In FUV about 50\% of the sources showed a harder when
brighter trend. The remaining sources too showed a harder when brighter
behaviour but the correlation is moderate. The weighted and un-weighted
linear least squares fit show similar trend for most of the sources, with the largest
mismatch seen in Mrk 478. In the NUV band there is moderate
correlation between $\alpha$ and the flux with a trend for a harder
when brighter behaviour. Here too, large discrepancy between weighted and un-weighted
linear least squares fits is seen in sources such as Mrk 478, NGC 4593 and 3C 390.3. 
These results to a large extent agree with the
analysis of the UV continuum emission in AGN by Sakata 
et al. (2011) who too found a bluer when brighter trend in their sample. Similar
conclusion was also arrived at by Wilhite et al. (2005) and
Vanden Berk et al. (2004) in the optical band. Our
results for a majority of the sources are also consistent with the observations of  Paltani \& Walter (1996) who
found that the UV spectra of Seyfert galaxies becomes flatter with increased brightness 
of the sources. To explain these observations, Paltani \& Walter (1996) proposed the two component
model. According to this model, the observed flux is a superposition of two distinct spectral
components, with constant spectral shapes. One component is flux variable
while the other one is stable and the observed continuum variation is driven by the amplitude
of the varying component.
For some sources in our sample such as NGC 3783, NGC 4151, NGC 4593 and Fairall 9, 
we in fact observed a constancy of the spectral index with increasing flux, however, only beyond
certain flux levels in them. This points to the complex nature of 
UV flux variations in AGN (cf. Paltani \& Walter (1996).  
The mean values of $\alpha$ for the sources in SWP and LWP are given in Table 2. Also, given in
the same table are the mean $\alpha$ values reported by Paltani \& Walter (1996) 
estimated using IUE spectra covering the wavelength range of 1150 $-$ 3200 \AA.
For the sources that are in common between this study and that of Paltani \& Walter (1996)
the mean $\alpha$ values are similar, though the data analysed here is much more
than that of Paltani \& Walter (1996).

\section{Lag between different wavebands}
\begin{figure}
\centering
\hspace*{-0.2cm}\includegraphics[scale=0.5]{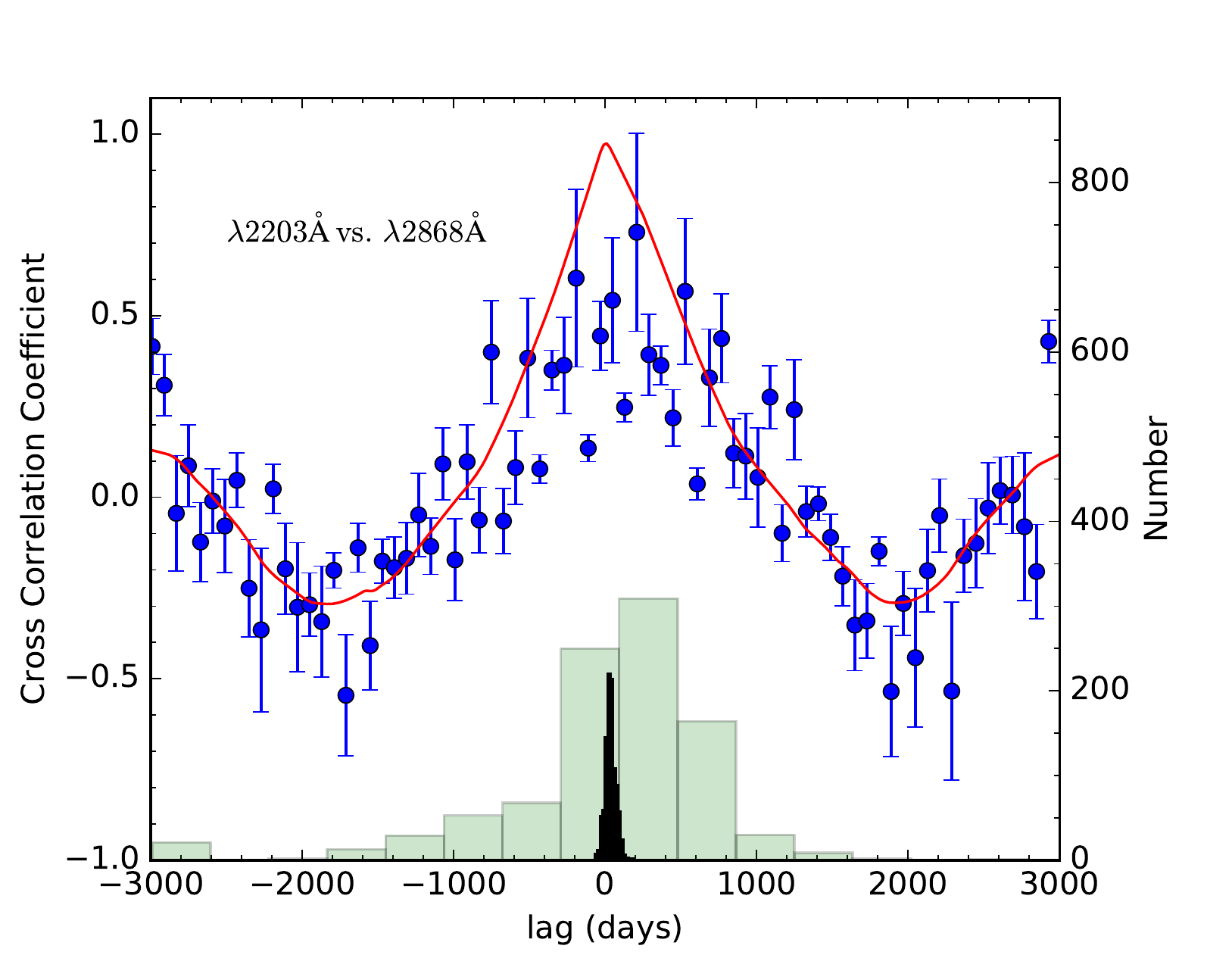}
\caption{Cross correlation analysis between the light curves at 2203 
\AA  ~and 
2868 \AA ~for the source  Fairall 9, one of the sources in the sample. The 
red solid line is for ICCF and the blue filled circles are for the DCF. The green and
black histograms show the distribution of centroids for DCF and ICCF respectively.} \label{Fig:12}.
\end{figure}

To check for inter-band time lags we used the discrete correlation
function (DCF) technique of Edelson \& Krolik (1988). The cross-correlation
analysis was done between the light curves of the shortest and longest
wavelengths in both FUV and NUV for all the sources. We show in 
Fig.\ref{Fig:12} the results on one correlation analysis for the object Fairall 9 
carried out between the light curves at 2203 and 2868 \AA . ~Here, the filled
circles are those evaluated using the DCF method and the solid line is that
obtained using the interpolated cross correlation function (ICCF) described
in detail in Gaskell \& Sparke (1986) and 
Gaskell \& Peterson (1987).  To evaluate
the uncertainty in the derived lag, we followed a model independent 
Monte Carlo approach that incorporated both flux randomization (FR) and
random subset sampling (RSS) described in Peterson et al. (1998). For
each Monte Carlo iteration we found 
the lag using the centroid of the CCF  utilizing
all points within 60\% of the peak of the CCF in the case of DCF. However, for
ICCF the peak of the CCF was considered as a representation of the lag between
the light curves. This was repeated for 10,000 times and the distribution of the
CCF lags were obtained for both DCF and ICCF methods. The mean of the 
distributions were taken to represent the lag between the light curves and
the spread in the distributions was used to estimate the error in the lag. The distributions
obtained using both using DCF (green histogram) and ICCF (black 
histogram) are given in Fig. \ref{Fig:12} for the source Fairall 9. We found no 
noticeable time lag between flux variations in NUV and FUV bands, though the 
flux variations between different NUV and FUV bands were correlated.
This analysis repeated for all the sources studied here, yielded no
measurable lags in any of them.
\section{Conclusion}
In the present work, we report the variability of fourteen 
Seyfert galaxies in
the UV band using data from IUE acquired over a period of about 17 years.
The flux values for the sources studied here  in different NUV and FUV bands
were taken from  Dunn et al. (2006). Various analysis were performed to characterize the
flux variability of the sources. The summary of the work is
given below

\begin{enumerate}
\item All sources were  found to show flux variations in the UV band. No statistically
significant difference in the amplitude of flux variations between shorter and longer
wavelengths was noticed.
\item No time lag between flux variations in different NUV and FUV bands was
observed
\item We found a mild negative correlation of variability with bolometric
luminosity with high luminous sources showing low variability than
their less luminous counterparts. Also,  a hint for a positive correlation 
is found between variability and black hole mass. These results are consistent 
with what is known in literature.
\item Majority of source showed a bluer when brighter trend in the FUV data,
however, such trend if any in NUV band is seen only in a minority of the 
sources that too moderately. Some sources showed a hardening of the spectrum
with flux, however, the spectrum remained non-variable beyond certain flux level. The 
observed spectral variations are thus complex.
\end{enumerate}

\section*{Acknowledgement}
We thank the anonymous referee for his/her critical comments that helped
to improve the manuscript



\begin{thebibliography}{99} 

\bibitem{2015PASP..127...67B} Bentz, M. C., Katz, S., 2015, PASP, 2015, 127, 67

\bibitem{1979PASP...91..589B} Bessel M.S., 1979, PASP, 91, 589

\bibitem{1989ApJ...345..245C} Cardelli, J.~A., Clayton, G.~C., \& Mathis, J.~S.\ 1989, ApJ, 345, 245 

\bibitem{2006PASP..118..572D} Dunn, J.~P., Jackson, B., Deo, R.~P., et al.\ 2006, PASP, 118, 572 

\bibitem{1995A&A..293..293A} D. Alloin et al.\ 1995, A\&A, 293, 293A

\bibitem{1992ApJ...401..516E} Edelson, R.\ 1992, ApJ, 401, 516 


\bibitem{1991ApJ...372L...9E} Edelson, R.~A., Saken, J., Pike, G., et al.\ 1991, ApJL, 372, L9 

\bibitem{1988ApJ...333..646E} Edelson, R.~A., \& Krolik, J.~H.\ 1988, ApJ, 333, 646 

\bibitem{2002apa..book.....F} Frank, J., King, A., \& Raine, D.~J.\ 2002, Accretion Power in Astrophysics, by Juhan Frank and Andrew King and Derek Raine, pp.~398.~ISBN 0521620538.~Cambridge, UK: Cambridge University Press, February 2002., 398 

\bibitem{1987ApJS...65....1G} Gaskell C. M., Peterson B. M., 1987, ApJS, 65, 1

\bibitem{1986ApJ...305..175G} Gaskell C. M., Sparke L. S., 1986, ApJ, 305, 175

\bibitem{1999MNRAS.306..637G} Giveon U., Maoz D, Kapsi S., Netzer H., Smith P.S., 1999, MNRAS, 306, 637

\bibitem{1997Ap&SS.248..261G} Greenhill L. J., Gwinn, C. R., 1997, Ap\&SS, 248, 261

\bibitem{2014ApJ...788...10L} Lohfink, A.~M., Reynolds, C.~S., Vasudevan, R., Mushotzky, R.~F., \& Miller, N.~A.\ 2014, Apj, 788, 10 

\bibitem{1969Natur.223..690L} Lynden-Bell, D.\ 1969, Nature, 223, 690

\bibitem{2013A&A...560A.104M} Meusinge H., Weiss V., 2013, A\&A, 560, A104

\bibitem{1994A&A...291...74P} Paltani, S., \& Courvoisier, T.~J.-L.\ 1994, 
A\&A, 291, 74 

\bibitem{1996A&A...312...55P} Paltani S., Walter R., 1996, A\&A, 312, 55

\bibitem{1997A&A...323..717P} Paltani S., Courvoisier T., 1997, A\&A, 323, 717

\bibitem{1998PASP..110..660P} Bradley M. Peterson, Ignaz Wanders, Keith Horne, Stefan Collier, 
Tal Alexander, Shai Kaspi and Dan Maoz

\bibitem{2017ApJ...842...96R} Rakshit S., Stalin C.S., 2017, ApJ, 842, 96

\bibitem{1984ARA&A..22..471R} Rees, M.~J.\ 1984, ARAA, 22, 471 

\bibitem{1997A&A..327..72R} Rodriguez-Pascual et al., A\&A, 72, 327

\bibitem{2011ApJ...731...50S} Sakata, Y., Morokuma, T., Minezaki, T., et al.\ 2011, ApJ, 731, 50 

\bibitem{2011ApJ...737..103S} Schlafly, E.~F., \& Finkbeiner, D.~P.\ 2011, ApJ, 737, 103 

\bibitem{2004JApA...25....1S} Stalin, C.~S., Gopal Krishna, Sagar, R., \& Wiita, P.~J.\ 2004, Journal of Astrophysics and Astronomy, 25, 1 

\bibitem{1997ARA&A..35..445U} Ulrich, M.-H., Maraschi, L., \& Urry, C.~M.\ 1997, ARAA, 35, 445 

\bibitem{2004ApJ...601..692V} Daniel E. Vanden Berk, Brian C. Wilhite, Richard G. Kron, Scott F. Anderson, Robert J. Brunner, Patrick B. Hall, Željko Ivezić, Gordon T. Richards, Donald P. Schneider, Donald G. York, 2004, ApJ, 601, 692

\bibitem{2003MNRAS.345.1271V} Vaughan, S., Edelson, R., Warwick, R.~S., \& Uttley, P.\ 2003, MNRAS, 345, 1271 


\bibitem{1995ARA&A..33..163W} Wagner, S.~J., \& Witzel, A.\ 1995, ARAA, 33, 163 

\bibitem{2001ApJ...556.1010W} Wang, X. Y.; Dai, Z. G.; Lu, T. 2001 ApJ, 556, 1010

\bibitem{2011A&A...527A..15W} Welsh, B.~Y., Wheatley, J.~M., \& Neil, J.~D.\ 2011, A\&A, 527, A15 

\bibitem{1998ApJ...509..118W} Welsh, W.~F., Peterson, B.~M., Koratkar, A.~P., \& Korista, K.~T.\ 1998, ApJ, 509, 118 

\bibitem{2005ApJ...633..638W} Wilhite, Brian C.; Vanden Berk, Daniel E.; Kron, Richard G.; Schneider, Donald P.; Pereyra, Nicholas; Brunner, Robert J.; Richards, Gordon T.; Brinkmann, Jonathan V. 2005, ApJ, 633, 638

\bibitem{2008MNRAS.383.1232W} Wilhite, Brian C.; Brunner, Robert J.; Grier, Catherine J.; Schneider, Donald P.; vanden Berk, Daniel E., 2008, MNRAS, 383, 1232

\bibitem{2007MNRAS.375..989W} Wold M., Brotherton M.S., Shang Z., 2007, MNRAS, 375, 989

\bibitem{2017MNRAS.464.2203Z} Zhang Xue-Guang, Feng, L, 2017, MNRAS, 464, 2203

\bibitem{2012ApJ...758..104Z} Zuo, Wenwen; Wu, Xue-Bing; Liu, Yi-Qing; Jiao, Cheng-Liang, 2012, ApJ, 758, 104
























\end{thebibliography}
\end{document}